\documentclass[notitlepage,aps,epsfig,showpacs,floats,amssymb,amsmath,floatfix,groupedaddress,superscriptaddress]{revtex4-1}
\usepackage{amsfonts,amssymb,stmaryrd,latexsym,amsmath,braket}
\usepackage{graphicx,subfigure}
\usepackage{comment}
\usepackage{times}
\usepackage{slashed}
\usepackage{bm}
\usepackage{braket}
\usepackage{appendix}
\usepackage{enumitem}
\usepackage{multirow}
\usepackage{array}

\usepackage[colorlinks=true,backref=true,linktocpage=true,citecolor=blue,urlcolor=blue,linkcolor=blue,pdfpagemode=UseOutlines]{hyperref}
\usepackage[dvipsnames]{xcolor}

\def\bnabla{\mbox{\boldmath $\nabla $}}

\begin{document}
\title{Effect of charge regulation on the screening properties of zwitterionic macroion solutions}
\author{Rashmi Kandari}
\affiliation{Department of Physics, Indian Institute of Technology, Jodhpur 342037, India}

\author{Rudolf Podgornik}
\email{Deceased}
\affiliation{School of Physical Sciences and Kavli Institute for Theoretical Sciences, University of Chinese Academy of Sciences, Beijing 100049, China}
\affiliation{Wenzhou Institute, University of Chinese Academy of Sciences, Wenzhou, Zhejiang 325000, China}
\affiliation{CAS Key Laboratory of Soft Matter Physics, Institute of Physics, Chinese Academy of Sciences, Beijing 100190, China}
\affiliation{Department of Physics, Faculty of Mathematics and Physics, University of Ljubljana, 1000 Ljubljana, Slovenia}

\author{Sunita Kumari}
\email{sunita@iitj.ac.in}
\affiliation{Department of Physics, Indian Institute of Technology, Jodhpur 342037, India}

\begin{abstract}
Precisely controlling the surface charge of zwitterionic macromolecules is crucial for tailoring their properties and optimizing them for specific applications. Here, we present a generalized calculation scheme for determining the screening length in solutions containing zwitterionic macroions, where the charge of the macroion is controlled by the electrolyte solution. This scheme bypasses the need to solve the Poisson-Boltzmann equation by expressing the inverse screening parameter in terms of the derivative of pressure with respect to chemical potential. The scheme reveals that the screening length has two components: one related to the Debye length with effective charges, and another stemming from the macroion surface dissociation equilibrium, which exhibits a ``screening resonance" behavior. Furthermore, we find that the nonuniform surface charge distribution induced by pH-responsive charge regulation strongly affects the screening behavior. The charge regulation properties of macroions, even in dilute solutions, are a key factor in the screening of electrostatic interactions, offering insights into complex biological and nanomaterial systems.
\end{abstract}

\date{\today}
\maketitle
\section{Introduction}
Electrostatic interactions are a fundamental component of (macro)molecular forces that span a wide range of scales, from colloid\cite{Isr11} to nanoscale domains\cite{French2010}. They play a particularly significant role in the biological and bio-molecular context, making a fundamental contribution to our understanding of nucleic acids\cite{bloomfield2000nucleic}, lipid membranes\cite{Cev18} and proteins\cite{Zhou2018}. The mean-field formulation of Coulomb fluid statistical mechanics\cite{Saf18} established electrostatic screening as a fundamental property of electrolyte solutions, standardly invoked in the description  of the macromolecular electrostatics\cite{Muthu2023}. The screening effect, determined by the screening length, influences how far the electrostatic effect of a charge can extend. This effect impacts various phenomena, including colloidal stability\cite{Yu2022,Kumar2017,SanchezFernandez2021}, DNA behavior\cite{Raspaud1998, Seo2019} and more\cite{Feng2019,Avni2022,Avni20A,Artemov2022,Balos2020}. For dilute electrolyte solution the most commonly used approach to the screening effect is the well known Debye H\"uckel (DH) theory. In the limit of large surface separation and correspondingly low surface charge density, the decay length of the double layer surface forces, as represented by the Poisson Boltzmann (PB), is indeed equivalent to the screening length\cite{breg10}. In concentrated electrolytes the charge density leads to an oscillatory decay of the potential, known as ``underscreening" or ``overscreening", where the observed screening length is larger than that predicted by the DH model\cite{Smith2016,gaddam2019electrostatic,lee2017scaling,adar2019screening}. In recent years extensive efforts have been made to explain experimentally observed screening anomalies, although these have been limited to those involving the constant surface charge (or constant surface potential) assumption\cite{martin1988sum,kirkwood1936statistical,kirkwood1954statistical,outhwaite1969extension,outhwaite1969extension,waisman1972mean,rasaiah1972calculations,hall1991modification,attard1993asymptotic,leote1994decay,caccamo1996integral,lee1996density,lee1997charge,varela2003exact,janevcek2009effective,girotto2017lattice,kjellander2018focus,Kjellander2019,kjellander2019intimate,yang2023solvent,lee1997charge,Kumari2022}.

\noindent Contrary to the fixed charge assumption, many zwitterionic macromolecules (dressed with anionic and cationic groups while remaining net-neutrally charged), such as PEGylated nanoparticles\cite{otsuka2003}, proteins bearing pairs of acidic and basic residues\cite{Isr11}, lipid bilayers\cite{marsh2006lipid} can dynamically exchange their ions in response to their local environment (such as, pH, ionic strength or even presence of other charged particles)\cite{Muthu2023, 1975regulation, Levin2023, Curk22}. This charge controlling mechanism modulates their electrostatic interactions, necessitating the concept of ``charge regulation" (CR)\cite{Muthu2023,ninham1971}. The ubiquitous nature of the CR mechanism has enabled the development of macromolecules for various applications, including therapeutic drug delivery\cite{otsuka2003}, anti-fouling surfaces\cite{Zheng2017,Baggerman2019}, and stimuli-responsive smart materials with diverse functionalities\cite{Netz2002,Borukhov2000,Kumar2012}.

\noindent Although the effects of charge-regulated electrostatics have been investigated in various scenarios\cite{1975regulation, 1976surface,, von1999,long2012, 2014field, 2015charge, 2015surface, 1996kin, 2004el,Landsgesell2019,levin2002electrostatic,Levin2023,Curk22}, the screening properties have not received proportionate attention.  The cell model, while useful for some initial estimations, doesn't account for the collective behaviour of macroions\cite{MaartenBiesheuvel2006,Boon2012}. In charged patchy surfaces, the interplay between CR and electrostatic screening results in nonlinear ionic screening, highlighting the intricate relationships between charge distribution, screening, and overall system behavior\cite{boon11}. The self-regulation of a mobile macroions modifies the charge distribution around it, leading to non-monotonic screening lengths and a buffering effect for small ions in the solution\cite{tomar17}. On the other hand, ``surface-specific interactions", or ``sticky length effects," significantly alter the effective screening length, indicating that the behavior of ions near surfaces is not simply a reflection of bulk solution properties\cite{tomar17, AVNI201970}. Within the framework of mean field(MF) theory, mobile macroions are treated as point-like particles in dilute solution, and it was shown that symmetric macroions exhibit a surprising behavior where they are most screened when half-filled, as opposed to when their sites are fully occupied\cite{Avni2018}. Furthermore, increasing the concentration of acceptor macroions(acquire ions from solution) in a solution leads to a transition where ions move from being primarily in the bulk to being mostly adsorbed on the macro-ions. This transition causes a non-monotonic dependence of the screening length on the macroion bulk concentration\cite{Avni2018}. However, this work did not consider the interactions between charged sites and the overall bulk charge of the macroion. Recently, based on the Frumkin Fowler Guggenheim (FFG) isotherm model\cite{koopal, mat01,rui23,sunita2024}, we have developed a calculation scheme to obtain the screening length in liquids composed of a single ion species and its counterionic macroions, which are subjected to CR by the former.\cite{sunita2024}. The screening parameter, which is the sum of two components: a Debye screening term and a term originating from charge fluctuations during the CR process. The Debye term accounts for the screening of charges due to counterions and macroions. While the second term captures the effects of fluctuations in charge distribution on the macroion surface that exhibit a rich and complex landscape of resonance screening phenomena\cite{sunita2024}. 

In general, the screening effect is determined by a wide spectrum of physicochemical parameters including ionic strength, size and valency of solute ions or the pH of the solution\cite{Lund2005,podgornik2018,Levin2023}. Unfortunately, no purely theoretical approach has been reported that covers so many variables in a single study. To overcome these hurdles, it is necessary to develop a fundamental understanding on the coupling of CR and screening behavior involving all system variables, in particular with macromolecules coated with zwitterionic surface dissociable groups. 

In this work, we focus on exploring the screening properties of charge regulated zwitterionic macroion in electrolyte solution. We will apply the FFG approach to model fluctuating charge surfaces as described in previous works\cite{sunita2024, rui23, koopal, mat01}. The paper is organized as follows: In the next Sec. \ref{system}, we introduce our system consisting of spherical macroions immersed in a univalent salt solution and free-energy and inverse screening length calculation procedure. The results and discussions are presented in Sec. \ref{result}. The paper ends with the final concluding remarks in the last Sec. \ref{final}.

\begin{figure}[!htp]
    \centering
    \includegraphics[width=.5\linewidth]{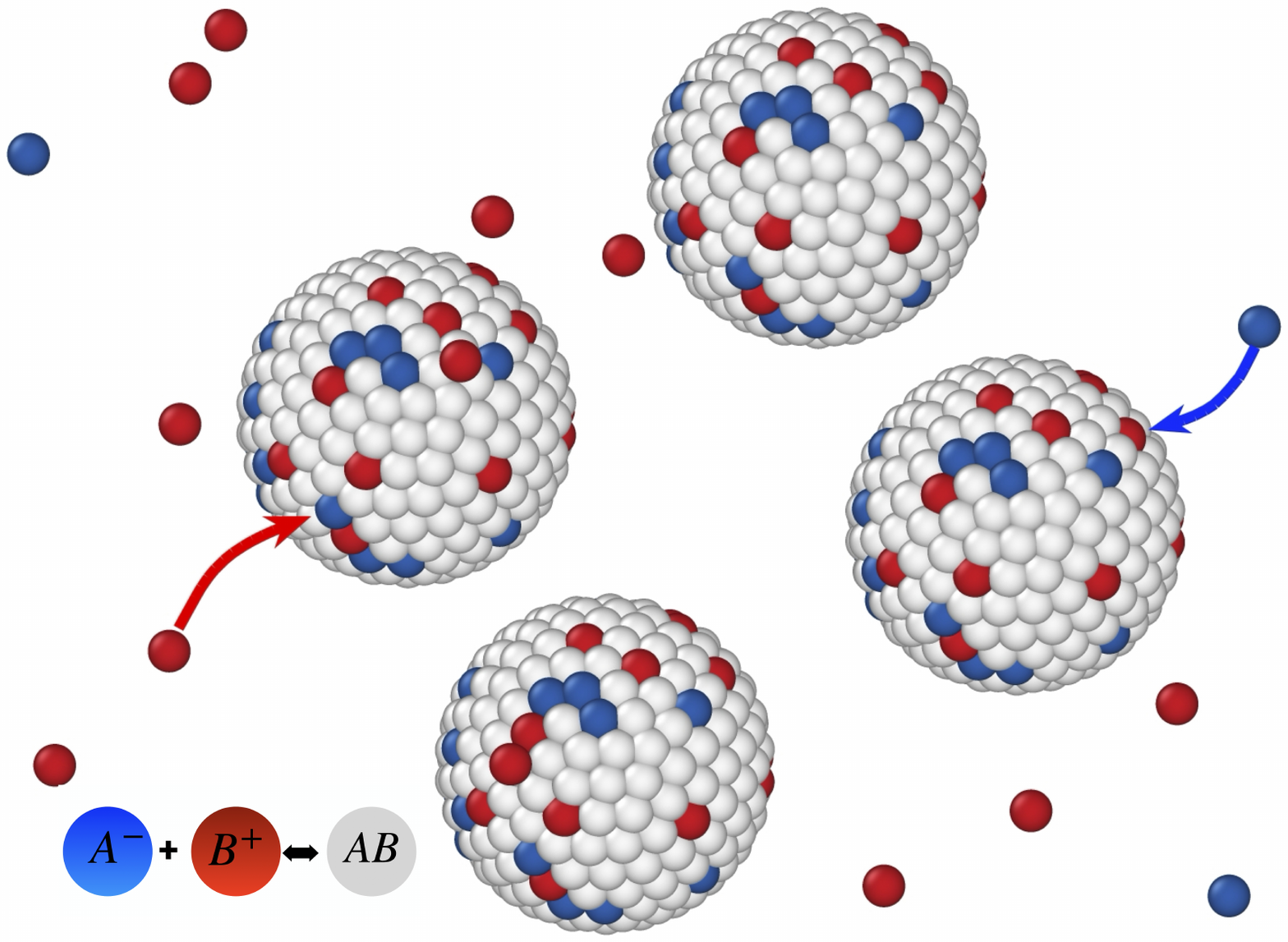}
    \caption{Model of the ionic solution containing macroions with surface dissociable groups undergoing a chemical reaction $A^{-}+ B^{+} \rightleftharpoons AB$. The macroscopic surface is embedded into a uni-valent electrolyte, composed of the electrolyte ions (small red and blue spheres) and macroions with dissociated/associated sites described by the same charge regulation model }
    \label{schematic}
\end{figure}

\section{Free energy calculation}
\label{system}
Fig. ~\ref{schematic} illustrates our model system composed of a solution ions (cation  ($``1"$) and anion ($``2"$)) and charge-regulated macroions with adsorbing/desorbing species $``1"$ and $``2"$. In addition, we assume that cations and anions in solution and dissociable groups on the macroion surface, are identical with respect to their chemical potentials, which correspond to similar ion species that can either be dissolved in solution or reside on the macroion. The description level is the same as for the macroion in the counterion solution and for the planar dissociable surface or single dissociable species\cite{sunita2024}. We recast free energy in terms of thermal units as\cite{Avni2018,AVNI201970},
\begin{eqnarray}
f[\phi_1, \phi_2, \phi_{3}; \phi'_1, \phi'_2] = f(\phi_1, \phi_2, \phi_{3}) +  \phi_{3} ~\tilde g(\phi'_1, \phi'_2).
\label{EQ8}
\end{eqnarray}
Here, $f(\phi_1, \phi_2, \phi_{3}; \phi'_1, \phi'_2)$ is the Poisson-Boltzmann (PB) electrostatic free energy density, which depends on the volume fractions of the salt cations ($\phi_{1}$), anions ($\phi_{2}$),  and macroions  $\phi_{3}$, as well as the volume fractions of the  cations $\phi'_1$ and anions $\phi'_2$  adsorbed onto the macroion surface. The last term ($\tilde g(\phi'_1, \phi'_2)$), refers to the charge regulated surface free energy density, dependent on the surface fraction of the same ions adsorbed to the surface consistent with previous work\cite{sunita2024, tomar17}. In Eq.~\ref{EQ8}, we assume the following ansatz for the bulk free energy for a charge-
regulated surface:
\begin{eqnarray}
f(\phi_1, \phi_2, \phi_{3}) &=&   ~{\phi_1} \left( \log{\phi_1} -1\right)  +  {\phi_2} \left( \log{\phi_2} -1\right) +  ~{\phi_{3}} \left( \log{\phi_{3}} -1\right),
\label{VO11abc12}
\end{eqnarray}
and for charge regulating  part we assume the ``Frumkin-Fowler-Guggenheim (FFG) isotherm model''\cite{har06, Koopal2020, Mozaffari_Majd2022} which implies,

\begin{eqnarray}
\tilde g(\phi'_1, \phi'_2) = - \sum_{i = 1,2} N_{i}\phi'_i \alpha_i + \sum_{i = 1,2} {N_{i}} \left( \phi'_i \log{\phi'_i} +  (1-\phi'_i )\log{(1-\phi'_i)}\right) + {\textstyle\frac12}\sum_{i = 1,2} \chi_{ij} \phi'_i\phi'_j,
\label{VO22deqwa}
\end{eqnarray}

Here, $N_{1,2} = n_{1,2}/N_3$ is the number of positive/negative dissociation sites on the macroion surface, $n_{1,2}$ and $N_3$ are the total number of cations, anions, and macroions, respectively. We assume that the interaction term is diagonal and takes the form, $\chi_{ij} = N_{i} \chi_i \delta_{ij}$. The term $\chi_{ij}$ represents pairwise interactions between
adsorbed surface sites and is of the general type as used in the Flory-Huggins theory\cite{mau41, paul}. $\chi \geq 0 (\chi \leq 0)$ refers to the strength of the attraction(repulsion) between adsorbed pairs. Our model is also valid in the absence of this pairwise interaction($\chi=0$). For simplicity, we consider that the salt ions and ionizable sites have identical sizes ($a$). The first three terms of the Eq.~\ref{VO22deqwa}, describe the adsorption mechanism with an energy cost $\alpha$ and an entropy given by the lattice gas theory. $\alpha$ term represents nothing other than the Langmuir adsorption isotherm and quantifies the interaction between macromolecules and their counter-ions. Actually, $\alpha$ and $\chi$ are connected with the equilibrium dissociation constant and its dependence on the dissociation strength. We now focus on the formulation of the CR problem for macroscopic surfaces by taking into account Legendre transform\cite{maggs2016} but implemented on macroions immersed in solution\cite{podgornik2018}. Given electrostatic energy as a function of the dielectric displacement vector ${{\bf D}} = - \varepsilon \nabla \psi$, the thermodynamic potential is modified as\cite{maggs2016,sunita2024},
\begin{eqnarray}
  {\cal F}[\phi_1, \phi_2, \phi_{3}, \phi'_1, \phi'_2, {\bf D} ] &=& \int_V\!\!\!d^3{\bf r} \Big( f(\phi_1, \phi_2, \phi_{3}) + \phi_{3} ~\tilde g(\phi'_1, \phi'_2) - \mu_1 {\phi_1} - \mu_2 {\phi_2} - \mu_3 {\phi_{3}} -  \phi_{3}\left(  {N_{1} \phi'_1} \mu'_1  +  
 {N_{2} \phi'_2} \mu'_2~\right)\Big)  \nonumber\\
 &&  + \int_V\!\!\!d^3{\bf r} \left( \frac{{\bf D}^2}{2\varepsilon} -
  \psi\left( \nabla\cdot {\bf D} - e_1\phi_1 - e_2 \phi_2 - {\phi_{3}} \left( e_1N_{1}\phi'_1 + e_2N_{2}\phi'_2 \right) \right)\right),
  \label{VO5}
\end{eqnarray}
with $e_i$ such that $e_{1,2}$ are the charges of the salt cations/anions and $e_{3}$ is the charge of the macroion resulting from CR $(e_1, e_2, e_3) \equiv   (e, -e, e\left( {N_{1} \phi'_1}  - {N_{2} \phi'_2}\right))$. Here, $\varepsilon = \epsilon\epsilon_0$, with $\epsilon$ and $\epsilon_0$ are the dielectric constant of the aqueous solution and vacuum, respectively. In Eq.~\ref{VO5}, the Poisson equation is introduced via a Lagrange multiplier, being the electrostatic potential $\psi$. In the general case, chemical potentials are simply displaced by electrochemical potentials such as, $ \mu_{1,2} \longrightarrow \mu_{1,2} \mp e \psi,$ $\mu'_{1,2} \longrightarrow  \mu'_{1,2} \mp e \psi$,  $\mu_3 \longrightarrow  \mu_3 + {N_{1} \phi'_1}(\mu'_1 - e \psi)  - {N_{2} \phi'_2}(\mu'_2 + e \psi),$ so that we end up with the chemical potentials of different species are as,
\begin{eqnarray}
\mu_{1,2} - e_{1,2} \psi &=& \frac{\partial f}{\partial \phi_{1,2}} \nonumber\\
\mu_{3} + \Big( \left( \mu'_1 - e_1\psi\right) N_{1} \phi'_1 + \left(\mu'_2 - e_2\psi\right) N_{2} \phi'_2 \Big) 
&=& \frac{\partial f}{\partial \phi_{3}} + \tilde g(\phi'_1, \phi'_2) \nonumber\\  N_{1,2} \phi_{3}\left( \mu'_{1,2}  - e_{1,2} \psi\right) &=&  \phi_{3} \frac{\partial \tilde g}{\partial \phi'_{1,2}}
\label{potential}
\end{eqnarray}
In the end, we also need to take into account the fact that the adsorbing/desorbing species is the same as the mobile ions and therefore: $\mu_{1,2} = \mu'_{1,2}$ plus that the mobile ions are a uni-univalent salt: $\mu_1 = \mu_2$. The general procedure for achieving a thermodynamic equilibrium state is to minimize the free energy functional, ${\cal F}[\phi_1, \phi_2, \phi_{3}, \phi'_1, \phi'_2, {\bf D} ]$. Minimization of Eq.~\ref{VO5} first with respect to $\textbf{D}$, and then potential $\psi$, gives Poisson equation,
\begin{eqnarray}
\bnabla\cdot {\bf D} - e_1\phi_1 - e_2 \phi_2 - e {\phi_{3}} \left( N_{1}\phi'_1 - N_{2}\phi'_2 \right) =0.
\label{defrho}
\end{eqnarray}
However, it would be preferred to write this free energy in the form of the electrostatic and chemical potentials to obtain the equivalent thermodynamic potential in the duplex form of the Eq.~\ref{VO5} such as,
\begin{eqnarray}
  {\cal G}[\mu_1, \mu_2, \mu_3, \mu'_1, \mu'_2, \psi] =   - \!\!\!\int_V\!\!\!d^3{\bf r}
  ~\Big( {\frac{1}{2}} {\varepsilon} (\bnabla\psi)^2 + p\Big(\mu_1 - e_1 \psi, \mu_2 - e_2 \psi, \mu_3 -\tilde p(\mu'_1 - e_1\psi , \mu_2 - e_2 \psi)\Big).
  \label{VO7Aa}
\end{eqnarray}
Let us introduce the Legendre transform method of the free energy or equivalently the osmotic pressure as a function of chemical potential as
\begin{eqnarray} \nonumber
&& p\Big(\mu_1 - e_1 \psi, \mu_2 - e_2 \psi, \mu_3,  \mu'_1 - e_1\psi, \mu'_2 - e_2\psi)\Big) = f(\phi_1, \phi_2, \phi_{3})- (\mu_1 -e_1 \psi) {\phi_1} - (\mu_2 -e_2 \psi){\phi_2} \\
&-& \big(\mu_3 + {N_{1} \phi'_1}(\mu'_1 - e_1 \psi)  + {N_{2} \phi'_2}(\mu'_2 - e_2 \psi) + (\tilde p(\mu'_1 - e_{1}\psi, \mu'_2 - e_{2}\psi)\big)\phi_{3}
  \label{VO7A2}
\end{eqnarray}

with
\begin{eqnarray}
\tilde p(\mu'_1 - e_{1}\psi, \mu'_2 - e_{2}\psi) =  \tilde g(\phi'_1, \phi'_2) -  {N_{1} \phi'_1} (\mu'_1  - e_{1}\psi) - {N_{2} \phi'_2} (\mu'_2  - e_{2}\psi),
\label{crm}
\end{eqnarray}

Eq.~\ref{VO7Aa} is very similar to the case of CR in the case of macroscopic bounding surfaces, but in that case, the CR contributes only to the boundary condition and not to the screening length. In the case of fixed charge, solution cations, anions, and macroions are different species with an electrochemical potential $\mu_1 - e_{1}\psi$ and $\mu_2 - e_{2}\psi$ and $\mu_3$. However, Eq.~\ref{crm}, shows that if the macroion charges are regulated, then what happens can be reproduced as a renormalization of $\mu_3$ to $\tilde g(\phi'_1, \phi'_2) -  {N_{1} \phi'_1} (\mu'_1  - e_{1}\psi) - {N_{2} \phi'_2} (\mu'_2 - e_{2}\psi)$. This renormalization of the fluctuating charge arises from the fact that the solution ions adsorbed onto the macroion surface causing the modulation of surface potential and charge states from $e_{i}N_{i}\phi_{3}$ to $e{\phi_{3}} \left( N_{1} \phi'_1 - N_{2}\phi'_2 \right)$. We now restrict ourselves to deriving the mean-field equations and then obtaining the screening parameter directly from the pressure derivative and the charge density relation, without actually solving the PB equation.


\subsection{Saddle-point (Poisson-Boltzmann) approximation and screening length calculation}
\label{saddle}
Now we derive the saddle-point equation (equivalently the PB equation) by invoking Legendre transformed free energy (Eq.~\ref{VO7Aa}). From here, the Euler-Lagrange equation takes the form,
\begin{eqnarray}
- \varepsilon \nabla^2 \psi =\rho(\psi) = - \frac{\partial }{\partial \psi}  p\Big(\mu_1 - e_1 \psi, \mu_2 - e_2 \psi, \mu_3 -\tilde p(\mu'_1 - e_1\psi , \mu_2' - e_2 \psi)\Big),
  \label{VO121abcd}
\end{eqnarray}
where the charge density is represented by $\rho(\psi) $. The derivative on the r.h.s by taking into account the Gibbs-Duhem relations. We follow the same steps that we took in our previous work\cite{sunita2024} on the single-site CR process and obtain the density as a function of the electrostatic potential and volume fractions, which gives us the PB Eq.~\ref{VO121abcd}. From here, expanding to the lowest order in terms of the electrostatic potential, one can derive that the final equation for the inverse screening length,
\begin{eqnarray}
 \tilde{\kappa}^2 = \frac{1}{\varepsilon} \frac{\partial \rho(\psi)}{\partial \psi} \bigg\rvert_{\psi = 0} =\frac{1}{\varepsilon}  \frac{\partial^2 p}{\partial \psi^2}   \sim  \frac{4\pi  l_B}{e^2}\Bigg(\underbrace{\sum_{i,j=1,2,3} e_i e_j \left( \frac{\partial^2 ~p}{\partial \mu_i\partial \mu_j}\right)_{e_i}}_{standard~screening~length} + \underbrace{\phi_{3}~\sum_{i,j=1,2} \tilde e_i \tilde e_j  \left(\frac{\partial^2 ~\tilde p}{\partial \mu'_i\partial \mu'_j}\right)_{\phi_i}}_{CR~contribution}\Bigg),
 \label{bfxjkwuyg}
\end{eqnarray}
with the first, intrinsic part containing all mobile species and their charges, while the last term is concerned with fluctuating charges. Note the summation over $i,j=1,2,3$ in the first part and over $i,j=1,2$ in the second. 
A huge simplification in these calculations is brought about by the ``indeterminacy relation" of the Legendre transform as explained in Ref\cite{Zia2009}. This relation states that
\begin{eqnarray}
\sum_{i,j=1,2,3}\frac{\partial^2 p}{\partial \mu_i\partial \mu_j} \frac{\partial^2 f}{\partial \phi_j \partial \phi_k} = \delta_{ik} \qquad {\rm and} \qquad \sum_{i,j=1,2}\frac{\partial^2 \tilde p}{\partial \mu'_i\partial \mu'_j} \frac{\partial^2 \tilde g}{N_{i} N_{j}\partial \phi'_j \partial \phi'_k} = \delta_{ik} 
\end{eqnarray}
The screening properties are therefore given directly by the second derivative of the free energy, {\sl i.e.}, in terms of the relevant response functions. The above simple form of the inverse screening length is completely consistent with the standard derivation that would proceed from the PB equation. It is straightforward to see that for the regular PB case, where $f$ is given by the ideal gas entropy, with all the terms stemming from the macroions absent, the above formula reduces exactly to the Debye screening length\cite{sunita2024}. The interpretation of the Eq.~\ref{bfxjkwuyg} form of the inverse square of the screening length is, therefore, composed of the Debye length evaluated with effective charges of the ions and macroions, plus a term corresponding to the macroion surface dissociation equilibrium. Omitting the common factor $e^2$ from Eq.~\ref{bfxjkwuyg}, the corresponding screening length then turns out as
\begin{eqnarray}
\frac{\tilde{\kappa}^2}{4\pi l_B}  = \kappa^2  &=& \phi_1 + \phi_2 + \phi_{3}\left( N_{1}\phi'_1 -  N_{2}\phi'_2\right)^2 + \phi_{3} \Bigg( \frac{N_{1}}{\frac{1}{\phi'_1}\frac{1}{(1-\phi'_1)} + {\chi_{1}}}  + \frac{N_{2}}{\frac{1}{\phi'_2}\frac{1}{(1-\phi'_2)} + {\chi_{2}}} \Bigg). ~~~~~~~~~~~~~~
 \label{fe}
\end{eqnarray}

Where $\kappa^2$ is rescaled by $4\pi l_B$. The $\kappa^2$ term explicitly depends on the number of solution ions, the number of dynamic surface sites, and the volume fractions of macromolecules and surface dissociation parameters. The dependence $\kappa^2$ indicates a complex functional dependence in contrast to the linear PB case. In Eq.~\ref{fe}, the contributions of solution ions and CR macroions are calculated by the initial three terms. The last term arises due to ion exchange between the solution and macromolecules.


\section{Result and Discussion}
\label{result}

A detailed look at the concentration dependence of the inverse screening length, in the limiting case of the electroneutrality clearly shows that the $\kappa^2$ is influenced by several factors, including the fraction of macroions, adsorption energy and the Flory–Huggins parameter $\chi$ (Eq.~\ref{finaleq}). In the following, we  focus on investigating how changes in system parameters affect different aspects of screening. Furthermore, since the salt is univalent $\phi_1 = \phi_2 = \phi$ and then because the identities of the dissociating ions and salt ions are assumed to be the same $\mu'_{1,2} = \mu_{1,2}$, allowing us to derive $\phi'_{1,2} = \phi'_{1,2}(\phi, \alpha_{1,2})$. Assuming furthermore a symmetric case corresponding to $\alpha_1 = \alpha_2$ and $\chi_1 = \chi_2 = \chi$, so that consequently $\phi'_{1} = \phi'_{2} = \phi'(\phi, \alpha, \chi)$, where

\begin{eqnarray}
  \log{\phi} = -\alpha_{1,2} +  \log{\frac{\phi'_{1,2}}{1-\phi'_{1,2}}}+\chi_{1,2} \phi'_{1,2}
\end{eqnarray}

which yields $\phi'_{1,2} = \phi'_{1,2}(\phi)$ that can be obtained implicitly from
\begin{eqnarray}
\phi'_{1,2}(\phi) =  (1 + \phi^{-1}e^{-\alpha_{1,2} + \chi_{1,2} \phi'_{1,2}(\phi)})^{-1}
\end{eqnarray}

Electroneutrality in the bulk further demands that
$ \phi_2 + {\phi_{3}} N_{2} \phi'_2  = \phi_1 + {\phi_{3}} N_{1}\phi'_1$, {\sl i.e.}, the total concentrations of the positive and negative mobile charges should be the same. This is trivially satisfied for the symmetric case, $\phi_1 = \phi_2 = \phi$ and $\phi'_{1} = \phi'_{2}$. Therefore, in the symmetric case, we remain with the simplified form

 \begin{eqnarray}
 \kappa^2 \sim  2 \phi + \phi_{3} \left( N_{1} -  N_{2}\right)^2 \phi'(\phi, \alpha, \chi)^2 +  \phi_{3} \Big( \frac{N_{1}}{\frac{1}{\phi'(\phi, \alpha, \chi)(1-\phi'(\phi, \alpha, \chi))} + {\chi}}  + \frac{N_{2}}{\frac{1}{\phi'(\phi, \alpha, \chi) (1-\phi'(\phi, \alpha, \chi))} + {\chi}} \Big). ~~~~~~~~~~~~~~
 \label{finaleq}
\end{eqnarray}
In this case, the screening length can be calculated explicitly as $\kappa = \kappa(\phi, \phi_{3}, \chi, \alpha)$. Analyzing the different conditions to our Eq.~\ref{finaleq}, it is worth mentioning that in the absence of charge regulation (CR) $(\phi_{3} \rightarrow 0$), the original Debye screening length solution is recovered, 
\begin{eqnarray}
   \lim_{\phi_{3}\to 0}  \frac{\kappa^2}{2\phi} \rightarrow 1  
\end{eqnarray}


Now we turn our attention to pH dependence on screening behaviour. By considering the variable ionization of surface groups, we are effectively linking the screening length, $\kappa$ to the pH and ion concentration of the solution. Considering this Eq.~\ref{finaleq},
recast as,
\begin{eqnarray}
\kappa^2 = \phi_1 + \phi_2 + \phi_{3} \left( N_1 \phi_1'(\mathrm{pH}) - N_2 \phi_2'(\mathrm{pH}) \right)^2 + \phi_{3} \left( \frac{N_1}{\frac{1}{\phi_1'(\mathrm{pH}) (1 - \phi_1'(\mathrm{pH}))} + \chi_1} + \frac{N_2}{\frac{1}{\phi_2'(\mathrm{pH}) (1 - \phi_2'(\mathrm{pH}))} + \chi_2} \right)
\label{ph}
\end{eqnarray}

Thus, the surface site fraction $\phi'_{1,2}$  now acquires explicit pH dependence, expressed as $\phi'_{1,2} = \phi'_{1,2}(\mathrm{pH})$. Recall that the Henderson-Hasselbalch equation\cite{hh2001, radic2012}, applied to ionizable surface groups on macroions, couples $\alpha$ to pH, quantifying the dissociation of acidic and basic groups, given by, 
\begin{eqnarray}
\alpha = \frac{1}{1 + 10^{\pm (pK - \mathrm{pH})}}
\quad \Rightarrow \quad
\phi'(\mathrm{pH}) = \frac{1}{1 + 10^{\pm (\mathrm{pH} - pK)}}
\end{eqnarray}

The pH-dependent screening resonance is characterized by the condition: $ \frac{d^2 \kappa^2}{d(\mathrm{pH})^2} = 0$. This criterion identifies a critical pH value where the screening behavior becomes highly sensitive to pH changes. At these points, slight pH variations can lead to significant changes in the screening behavior, enabling the modulation of electrostatic interactions through pH adjustment. To refine the concept of pH-dependent screening resonance, it is important to consider that the local pH near charged surfaces differs from the bulk pH. This effect must be incorporated as follows:
\begin{eqnarray}
pH_{\mathrm{local}} = pH_{\mathrm{bulk}} - \frac{\psi e_0}{k_B T \ln(10)}
\end{eqnarray}
Where $\psi$ represents the electrostatic potential at the surface. Surface potential pH shift is particularly important in concentrated solutions or near highly charged surfaces, affecting the ionization degree  of surface groups. In addition to surface effects, bulk pH shifts can also arise due to the Donnan potential\cite{Levin2023}, This leads to a redistribution of ions across the boundary and a potential difference, causing the internal pH to deviate from the external (bulk) value. The Donnan-induced pH shift can be approximated by:

\[
\Delta pH = -\frac{e \psi_{\text{Donnan}}}{k_B T \ln(10)}
\]

The surface potential arises from local charge at an interface, while the Donnan potential results from overall charge imbalance across a semi-permeable barrier or within compartments. In rare cases with thick, uniformly charged regions, the surface and Donnan potentials may coincide, but they typically differ in magnitude and range due to charge non-uniformity or thin layers.
In this system, both effects may be present: the Donnan effect sets the overall pH difference across regions, while the surface potential further modulates the pH right at the macroion surface, directly impacting CR and screening.

\begin{figure}[!t]
\centering
\includegraphics[width=15cm,height=7cm]{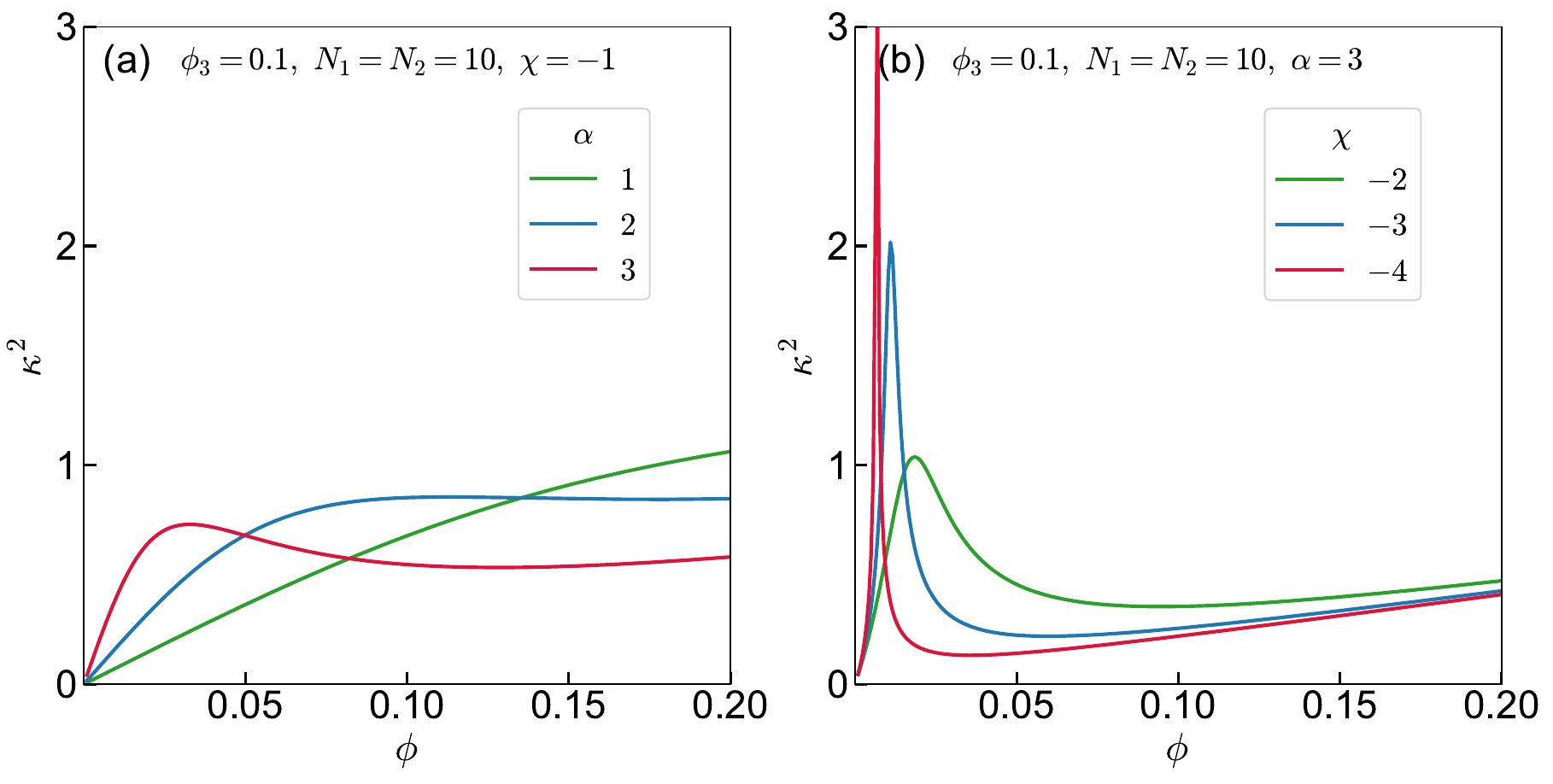}
\caption{(Color online) The square of inverse screening length ($\kappa^2$) as a function of salt concentration ($\phi$) as given in Eq.~\ref{finaleq}. In the left panel, the different curves correspond to different absorption energies ($\alpha=1,2,3$), while in the right panel, the results are shown for three different values of the interaction parameter between surface sites $\chi ( = -2, -3, -4)$. Note: for low $\phi$ and large negative values of $\chi$, $\kappa^2$ exhibits screening resonance.} 
\label{alpha}
\end{figure}

In Fig.~\ref{alpha}, we represent the dependence of the inverse screening length ($\kappa^2$) as a function of the solution concentration ($\phi$), interaction strength between ions and the macroion surface ($\alpha$), and interaction between similar macroion surface sites ($\chi$). The term $\alpha$ signifies the non-electrostatic part of the adsorption energy, and its positive value suggests a stronger attraction between the ions and the macroion surface, leading to adsorption. Conversely, a negative value ($\alpha < 0$) indicates a repulsive interaction, promoting dissociation. This occurs when electrostatic repulsion or steric hindrance dominates. The parameter $\chi$ is similar to the Flory–Huggins interaction parameter\cite{Teraoka2002}. It reflects how strongly the charge of a macroion is tuned by its bathing solution conditions and the resulting effects on the macroion's behavior. $\chi > 0$ indicates that the adsorbed ions prefer to stay  near each other (attraction), while a $\chi < 0$ indicates they prefer to be dissociated\cite{rui23,sunita2024}. The Langmuir model can be viewed as a special case of the Frumkin-Fowler-Guggenheim (FFG) model where $\chi=0$, implying there are no lateral interactions. For the fixed values of $\alpha $ and $\chi$, it can be seen from  Fig.~\ref{alpha} that $\kappa^2$ increases rapidly with increasing concentration until it reaches a maximum and shows a ``screening resonance."  This resonance behavior arises due to the maximum in charge fluctuations, as captured by the last term in Eq.~\ref{finaleq}, which represents the fluctuation contribution to the screening. As shown in Fig.~\ref{alpha}, at low salt concentrations (i.e., small $\phi$), the added salt ions are insufficient to effectively screen the electrostatic interactions between macroions. In this regime, charge-regulated macroions ($\phi_{3}$) themselves become effective screening agents. This occurs because their effective surface charge dynamically adjusts to the local ionic environment, enabling them to partially neutralize other macroions or surrounding ions.
As shown in Fig.~\ref{alpha}(a), if the strength of $\alpha$  is increased (meaning that solution ions are encouraged to bind to the macroion surface), the intensity of the screening peaks becomes  sharper and shifts towards the lower $\phi$ regime. On the other hand, when $\chi$ is negative and large in magnitude (this means that the dissociation of ionizable groups is promoted), the screening resonance peak becomes sharper and more pronounced, leading to a stronger screening effect, as evident in Fig.~\ref{alpha}(b). This unusually large screening effectively prevents the electric field from penetrating the macroion solution, meaning that the field is highly localized near the macroion surface and does not penetrate deep into the solution. This also confirms that even in solutions with low salt concentrations, charge modulated zwitterionic macroions can produce a screening effect. This is because the CR mechanism involves a redistribution of charges on the surface of the macroions, which compensates for the lack of sufficient free ions to screen the field. 

\noindent Fig.~\ref{alpha} illustrates that $\kappa^2$ tends to decrease monotonically with increasing salt concentration (large $\phi$), and the non-monotonic (fluctuating) behavior observed at lower concentrations disappears. This is because the dominant effect of increasing salt concentration becomes the screening of electrostatic interactions, which overshadows the impact of $\alpha$ and $\chi$ contributing to the nonmonotonic behavior, and the system behaves more predictably with monotonic screening; see Fig.~\ref{alpha}(b). This behavior remains valid even for large negative values of $\chi$. This situation is similar to that of a single-species macroion in a counterion solution\cite{sunita2024} and symmetric zwitterionic macroions in bulk concentrations\cite{Avni2018}.

\begin{figure}[!t]
\centering
\includegraphics[width=15cm,height=7cm]{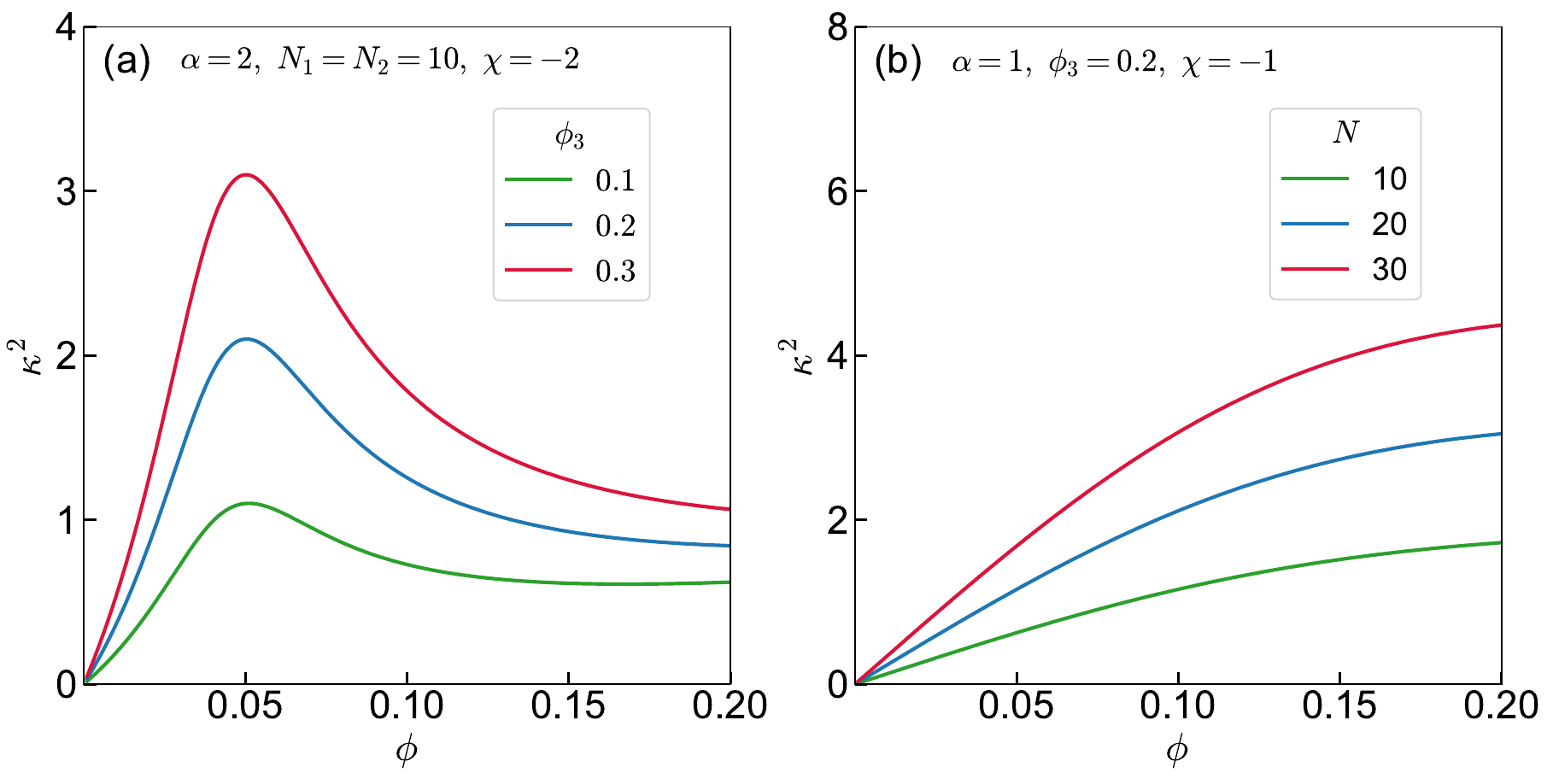}
\caption{(Color online) The square of inverse screening length ($\kappa^2$) as a function of salt concentration ($\phi$) as given in Eq.~\ref{finaleq}. In the left panel, the different curves correspond to different macroion concentrations($\phi_{3}$)), while in the right panel, the results are shown for three different values of surface sites $N (= N_1=N_2= 10, 20, 30)$.}
\label{kappa}
\end{figure}
\begin{figure}[!t]
\centering
\includegraphics[width=15cm,height=7cm]{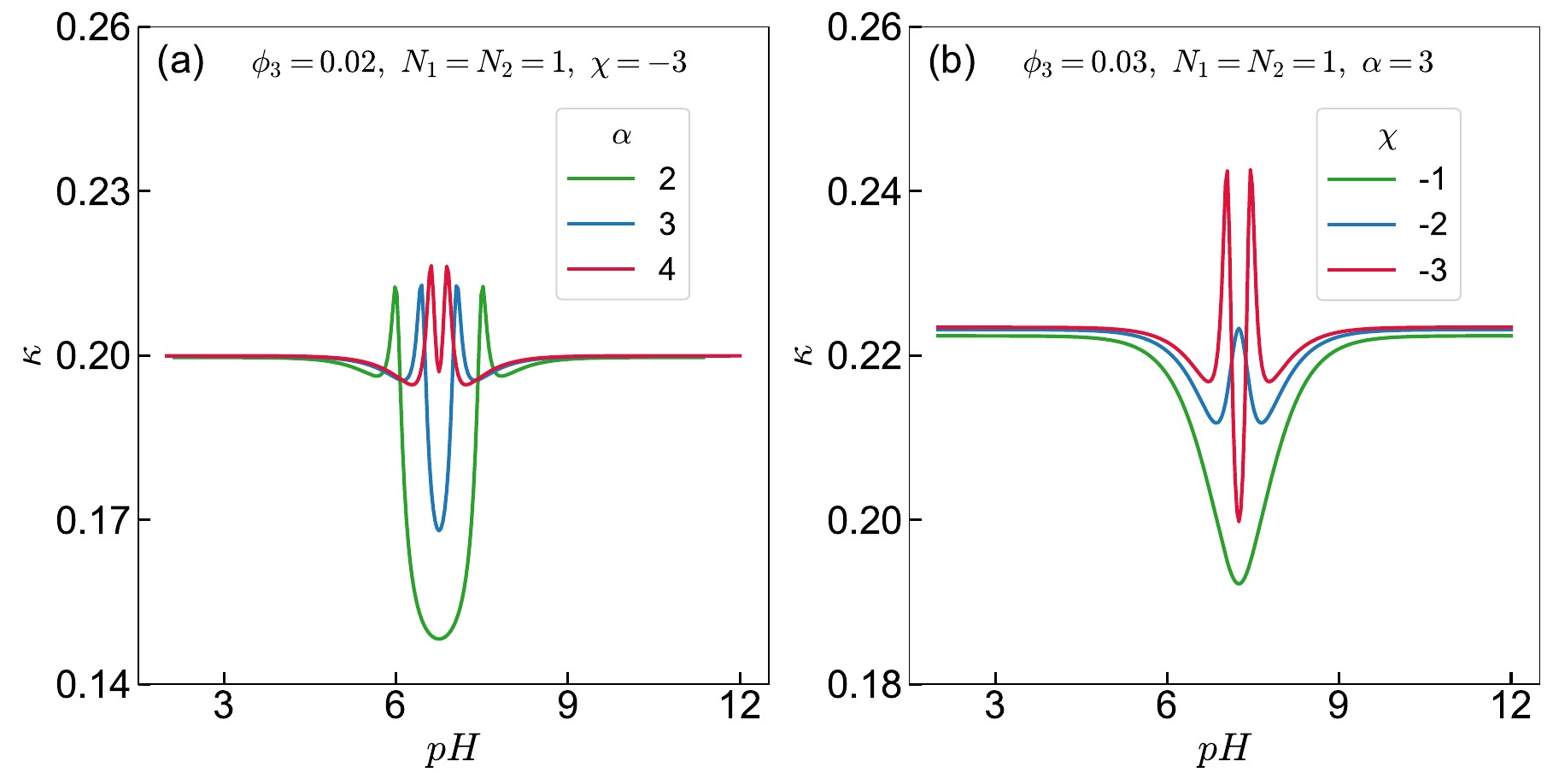}
\caption{(Color online) The  inverse screening length ($\kappa$) as a function of $pH$  as given in Eq.~\ref{ph} for (a) different curves correspond to adsorption energy($\alpha$)  with fixed pK:  4.5 (acidic) and 9.0 (basic). (b) $\kappa$ as a function of interaction parameter($\chi$) with fixed pK:  5.5 (acidic) and 9.0 (basic).}
\label{ph1}
\end{figure}
\begin{figure}[!t]
\centering
\includegraphics[width=15cm,height=7cm]{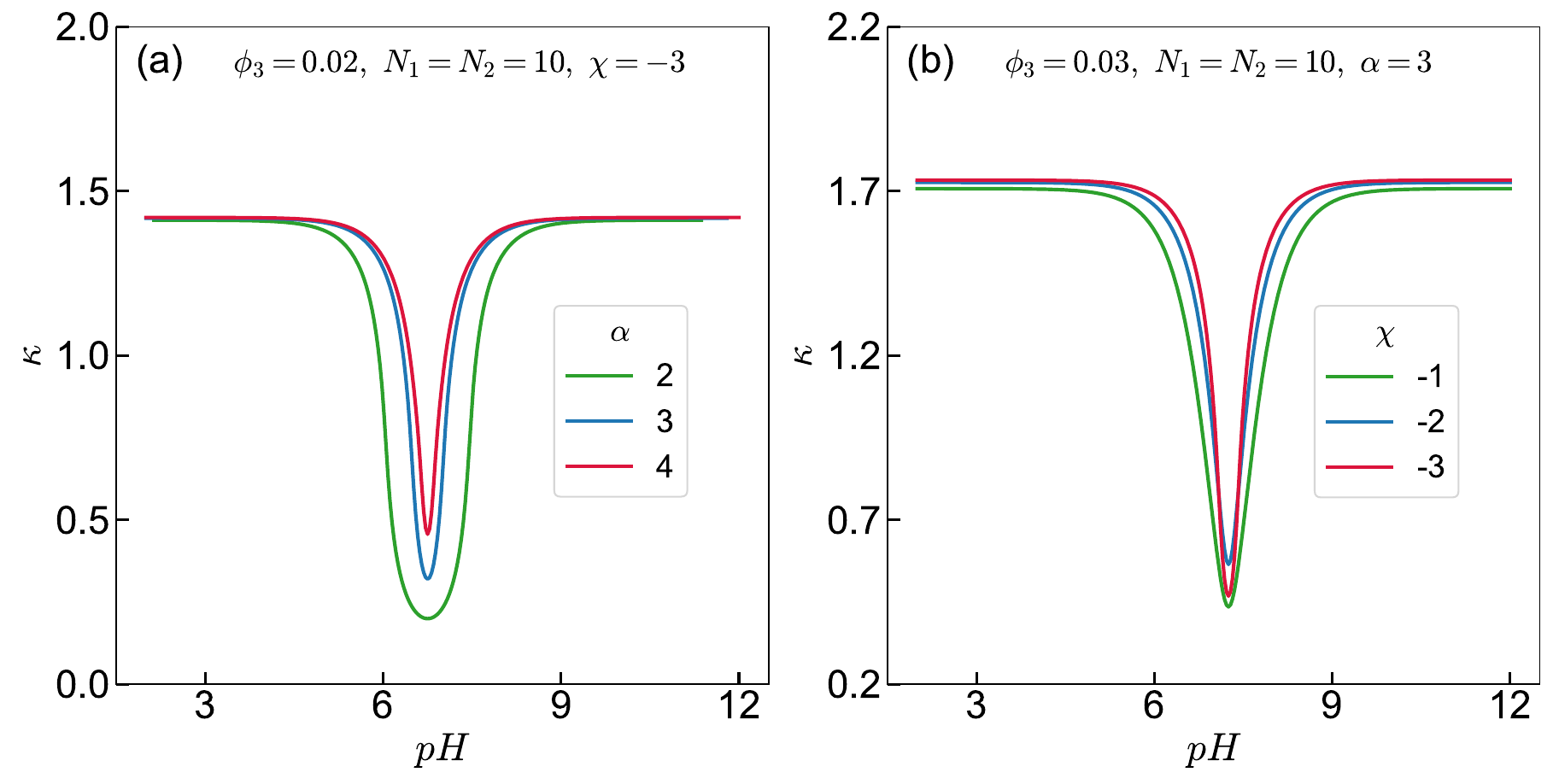}
\caption{(Color online) The  inverse screening length ($\kappa$) as a function of $pH$ as given in Eq.~\ref{ph} for (a) different curves correspond to adsorption energy($\alpha$)  with fixed pK:  4.5 (acidic) and 9.0 (basic). (b) $\kappa$ as a function of interaction parameter($\chi$) with fixed pK:  5.5 (acidic) and 9.0 (basic).}
\label{ph2}
\end{figure}

\noindent In Fig.~\ref{kappa}(a-b), we present the dependence of $\kappa^2$ on the macroion concentration ($\phi_{3}$) and the number of disassociated/associated acidic ($N_1$) and basic ($N_2$) groups. The resonance peak in the inverse screening length ($\kappa^2$), which represents the strength of electrostatic screening, is indeed strongly influenced by the $\phi_{3}$; see Fig.~\ref{kappa}(a). Furthermore, as  $\phi_{3}$ increases, the peak value of the $\kappa^2$ also tends to increase, indicating a stronger screening effect. For example, at fixed $\phi=0.05$,  the peak for $\phi_{3}= 0.3$ is more than three times that of $\phi_{3} =0.1$. The explanation for this is that as  $\phi_{3}$ increases, the macroions themselves begin to act as screening agents. 
\noindent The presence of fluctuating acidic($N_1$) and basic($N_2$) sites on macroions triggers screening in solutions\cite{Avni2018,tomar17,sunita2024} as displayed in Fig.~\ref{kappa}(b).  Note that, for $N_1=N_2=N$ = 30, $\kappa^2$ is significantly higher (fourfold) than that for $N_1=N_2=N$ = 10. This effect is significant because the fluctuating charges on the zwitterionic macroion surface can effectively compensate for the presence of free ions in the solution. The fluctuating charges on the macroion surface help to neutralize the charges of surrounding ions. When the solution contains more salt ions, the macroions respond by increasing their charge density and the number of fluctuating sites, thereby enhancing the $\kappa^2$. Further increasing the salt concentration does not significantly enhance the screening effect, as there are no more available sites for the ions to bind, leading to a saturation behavior in the screening effect. 

 Since CR depends on the local chemical environment particularly pH, the degree of surface group dissociation, and thus the effective surface charge, changes with pH. This introduces an additional mechanism by which pH modulates the screening length, beyond the effects of ion concentration alone. Therefore we further investigate how pH modulates the screening length. The variation of $\kappa$ as a function of pH as given in Eq.~\ref{ph}, for three representative values of $\alpha$ and $\chi$ are shown in Fig.~\ref{ph1}, respectively. Note that the $\kappa$ profile exhibits non-monotonic behavior as the $pH \rightarrow 7$, particularly when the system has a low density of acidic and basic sites for all three values of $\alpha(=2,3,4)$:(Fig.~\ref{ph1} (a)) and  $\chi(=-1,-2,-3)$: (Fig.~\ref{ph1} (b)). As depicted in Fig.~\ref{ph1}, the two minima corresponding to the protonation and deprotonation of acidic and basic residues on the macroions are pH dependent, which directly affects the overall charge of the macroion and hence the screening effect. The relationship between adsorption strength between macroion and ion ($\alpha$), dissociation strength between surface charged sites ($\chi$), and $\kappa$ is not always straightforward and exhibit complex, non-monotonic behavior when $\alpha$ and $\chi$ are changes with pH. However, the assumption of a single acidic/basic dissociation site ($N = 1$) for macroions is actually an oversimplification. In reality they usually have multiple absorption/dissociation sites on their surface. As the number of surface sites increases $N \geq 1$, the two minima merge together and become a single minimum, see Fig.~\ref{ph2}. It can be seen that $\kappa$ is clearly minimum when pH $= 7$ corresponds to weak electrostatic interactions. $\kappa$ is maximum when the total charge changes the most with pH (i.e. if the solution is acidic($pH \rightarrow 1$) or basic(pH $\rightarrow 12$)). With more dissociable charge sites, the relative effect of individual ion fluctuations becomes smaller, making the overall charge fluctuations less important; therefore, we see the saturation behaviour showing symmetry on both sides, whether acidic or basic. Notably, the $\kappa$ curve shows an inflection point near neutral to mildly basic or acidic value of pH ($5 \leq$ pH $\leq 9$), marking the shift from  saturation to rapid ionization, where further pH increases cause smaller changes in charge density as shown in Figs.~\ref{ph1} and \ref{ph2}. This intermediate pH range represents peak sensitivity in screening length and efficiency, highlighting a critical regime for pH-dependent electrostatics.  constant charge assumption, the effective screening length nonlinearly decreases with increasing pH\cite{gun2014effects}.

 It is important to remark that both the screening length and the differential capacitance are intrinsically linked to the surface charge density, and often exhibit parallel trends as a function of pH\cite{Lund2005, podgornik2018,Stornes2021,Boi2021,Bakhshandeh2022,Colla2024,Bakhshandeh2024,Bossa2024}. Consistent with this, our pH analysis shows similar behavior as that found for differential capacitance as a function of pH for charge regulated nanoparticle\cite{Stornes2021} and proteins\cite{Lund2005}, and for zwitterionic planar electrodes\cite{Bossa2024} in solution. The surface charge density as a function of pH has been reported for a system of colloidal and nanoparticle coated with  acidic groups\cite{Colla2024,Bakhshandeh2022,Bakhshandeh2024}. Furthermore, it was shown that for very high pH, the acidic groups become completely deprotonated, however, for pH $\leq 2$, nearly all sites are protonated. The addition of an electrolyte reduces electrostatic interactions due to a ``screening effect," which makes it harder for protons to bind to a molecule, even at the same pH\cite{Colla2024,Bakhshandeh2022,Bakhshandeh2024}.

\section{Final Remarks}
\label{final}
Zwitterionic materials, characterized by having both positive and negative charges on the same molecule, offer a unique combination of properties. The surface chemistry of these materials is strongly shaped by the surrounding fluid environment, including the presence of other charged molecules through a charge regulation (CR) process. Understanding the interrelationship between CR and screening is essential for designing and controlling the behavior of materials in a wide range of applications, from materials science and nanotechnology to biology and medicine. In this work, we use the Frumkin Fowler Guggenheim (FFG) isotherm approach to model  charge-regulated zwitterionic macroions immersed in a monovalent salt solution. The analysis focuses on understanding the screening length in systems with charge-regulated macroions by leveraging the ``uncertainty relation" derived from the Legendre transform. The analysis incorporates the free energy contributions from both the ions in the solution and the CR phenomena occurring at the macroion's surface. Our results show that the pH significantly modulates the screening properties of charged surfaces, and these effects are often mirrored in the behavior of surface charge density and surface differential capacitance as reported in several Refs.\cite{Lund2005, podgornik2018,Stornes2021,Boi2021,Bakhshandeh2022,Colla2024,Bakhshandeh2024,Bossa2024}.
We have shown that the inverse screening length exhibit a qualitatively different, and more complex, non-monotonic dependence on the salt concentration that cannot be replicated by static charge models or predicted by linear Poisson-Boltzmann theory. The inverse screening parameter displays the ``screening resonance" as a function of a subset of the system parameters. Even at low salt concentrations, the macroion's own dissociable charges create this local screening effect and show resonance behaviour. However, the screening resonance prediction presented above has not yet been validated by any experimental spectra. The lack of observed resonance screening in experiments can stem from a combination of factors: the specific experimental conditions, the intricacies of charge regulation in the system, and the constraints of the detection methods used. It is important to carefully consider these factors in designing experiments and interpreting results in this area.


\section{Acknowledgment}
SK acknowledges the financial support from the Science and Engineering Research Board (SERB), Department of Science and Technology (DST), India (grant No. EEQ/2023/000676) and IIT Jodhpur for a research initiation grant (grant no. I/RIG/SNT/20240068).

\bibliographystyle{apsrev4-1}
\bibliography{Manuscript}

\end{document}